\documentclass{ws-ijmpa}
\usepackage{amsmath}
\usepackage{graphicx}
\usepackage{multirow}
\usepackage[super,compress]{cite}
\newcommand\Tstrut{\rule{0pt}{2.6ex}}

\begin{document}

\markboth{Recombination in YMG}{}

%
\catchline{}{}{}{}{}
%

\title{Recombination of H and He in Yang-Mills Gravity}

\author{Daniel Katz}

\address{Department of Physics \& Applied Physics, University of Massachusetts Lowell\\
1 University Avenue
Lowell, MA, 01854
United States\\
daniel\_katz1@student.uml.edu}

\maketitle

\begin{history}
\received{Day Month Year}
\revised{Day Month Year}
\end{history}

\begin{abstract}
We investigate some aspects of the thermal history of the early universe according to Yang-Mills Gravity (YMG); a gauge theory of gravity set in flat spacetime. Specifically, equations for the ionization fractions of hydrogen and singly ionized helium during the recombination epoch are deduced analytically and then solved numerically. By considering several approximations we find that the presence of primordial helium and its interaction with Lyman series photons has a much stronger effect on the overall free electron density in YMG than it does in the standard, General Relativity (GR) based, model. Compared to the standard model recombination happens over a much larger range of temperatures, although there is still a very sharp temperature of last scattering around $2000\ K$. The ionization history of the universe is not directly observable, but knowledge of it is necessary for CMB power spectrum calculations. Such calculations will provide another rigorous test of YMG and will be explored in detail in an upcoming paper.

\end{abstract}

\ccode{PACS numbers:}

\section{Introduction}
The prediction~\cite{abc} and subsequent discovery~\cite{penz} of the Cosmic Microwave Background (CMB) radiation is one of the most compelling pieces of evidence in favour of the Big Bang theory of cosmic origins. General Relativity (GR), combined as needed with electro- and thermodynamics, does more than an adequate job of explaining the majority of observed astrophysical phenomena such as gravitational lensing, periapsis precession, the existence and properties of the CMB, and the relative abundances of light elements in the universe. The first two are examples of what can be done with a static point-source solution to Einstein's field equations (\emph{i.e.} Schwarzschild's solution) while the latter are typically accounted for by a ``whole-universe" solution via the Robertson-Walker (RW) metric. As useful and accurate as GR is, however, there are many compelling reasons to look beyond it for a more complete description of gravity. Chief among these are the difficulties associated with combining GR and Quantum Mechanics~\cite{Rovelli}. There are also issues with GR both predicting the existence of singularities like black holes and the Big Bang and its own breakdown in treating them~\cite{Strominger}. Because of this, alternative theories of gravity (usually quantum in nature) abound in the literature. In this article we consider the early thermal history of the universe according to one of them; Yang-Mills Gravity (YMG). In particular we analyse the recombination epoch. The same physics of an expanding plasma involved in the standard GR based approach applies to our treatment. However, we will see that the presence of helium makes a much more significant contribution to the evolution of the free electron density than it does in GR. As we obtain and present numerical results on ionization fractions for hydrogen and helium we will compare them quantitatively with standard results and find that, while they agree on the order of magnitude there is considerable disagreement in the details. For each ionization history we will calculate a photon visibility function which gives the probability that a given photon last scattered at the specified temperature. The fractional ionizations are not measurable quantities themselves, but the visibility functions they dictate are needed to calculate the angular power spectrum of the CMB. Calculating this power spectrum requires first working out a general theory of small fluctuations in an otherwise homogeneous isotropic universe described by YMG. Such a formalism is the motivation for this work but is its self still a work in progress.\newline
\indent This is not meant to be a review article on YMG, but because it is largely unknown in the community we will expound its principles and some of its consequences here for the readers' convenience. More information can be found in Refs.~\citen{Hsu-06,Hsu-09,Hsu-11}. Unless otherwise specified we will be using units where $c = \hbar = 1$ throughout.\newline

\section{Yang-Mills Gravity}
YMG is a gauge theory set in flat (Minkowski) spacetime. As with all gauge theories its base is a group, which in our case is the group of all $4$-dimensional spacetime translations which we'll call $T_4$. Physics is already invariant under global $T_4$ transformations, so we impose upon it the further restriction of being invariant under \emph{local} infinitesimal translations. Explicitly, we want nothing to change physically under the transformation
\begin{equation}
x^\mu \rightarrow x^{\mu \prime} = x^\mu + \Lambda^\mu (x)
\label{eq:trans}
\end{equation}

\noindent where $\Lambda^\mu$ is infinitesimal and behaves as nicely as we need it to mathematically but is otherwise arbitrary. Obviously, the Lagrangians (and hence the equations of motion) for most systems are not going to be invariant under Eq.~\ref{eq:trans}. Using the formalism of Yang and Mills~\cite{ym} (for which this theory is named) we can impose almost whatever local symmetry we like at the expense of adding a gauge field to the system. To do so we begin by noting that the generators of $T_4$ are $i\partial_\mu$ and so $T_4$ is abelian, which makes dealing with it somewhat easier. On the other hand, its generators carry Lorentz indices so we cannot suppress them when expressing the $T_4$ gauge covariant derivative
\begin{equation}
\partial_{\mu} \rightarrow D_\mu = \partial_\mu + g\phi^\nu_\mu \partial_\nu,
\label{eq:cov}
\end{equation}

\noindent where $g$ is a coupling constant to be determined later and $\phi^{\mu \nu}$ is the (symmetric) tensor field associated with local $T_4$ symmetry. The gauge field is a tensor field in this case because the index of the group generators is its self a spacetime index and so the field needs to carry two such indices. So far it is not obvious that this theory describes gravity, although two hints emerge: $1)$ the covariant derivative is its own complex conjugate, so whatever force is mediated by $\phi^{\mu \nu}$ has only attractive modes; and $2)$ the quanta of a tensor field can have spin $2$, in agreement with general expectations for the graviton. For any Lagrangian we can make the replacement of Eq.~\ref{eq:cov} and obtain a locally $T_4$ invariant version of that Lagrangian. This can in turn be separated into the original Lagrangian plus a coupling term which describes how $\phi^{\mu \nu}$ interacts with other particles in the system. For this to be useful we'll need to know the dynamics of $\phi^{\mu \nu}$ by its self. We construct the simplest possible Lagrangian density for our gauge field as follows. The gauge field strength is represented by the commutator of the covariant derivative with its self:
\begin{eqnarray}
[D^\mu, D^\nu] &=& C^{\mu \nu \alpha}\partial_\alpha \nonumber \\
C^{\mu \nu \alpha} &=& J^{\mu \sigma}\partial_\sigma J^{\nu \alpha} - J^{\nu \sigma}\partial_\sigma J^{\mu \alpha} \nonumber \\
J^{\mu \nu} &=& \eta^{\mu \nu} + g\phi^{\mu \nu}.
\label{eq:str}
\end{eqnarray}

\noindent Here $C^{\mu \nu \alpha}$ is the field strength, $\eta^{\mu \nu}$ is the Minkowski metric, and $J^{\mu \nu}$ is a symbol which is often easier to work with than $\phi^{\mu \nu}$ directly, defined for convenience. In this paper we will stick with $\eta^{\mu \nu}$ as our background metric, but this is not required. Any geometrically flat metric tensor, \emph{i.e.} one for which the Reimann curvature tensor vanishes, would do as well. The metric chosen must correspond to zero curvature because flat spacetime is an assumption of YMG. If another flat metric, say $P^{\mu \nu}$, is chosen then the partial derivatives in eqns~\ref{eq:cov} and~\ref{eq:str} need to be replaced with covariant derivatives with respect to $P^{\mu \nu}$. \newline
\indent The simplest Lagrangian density we can construct for $\phi^{\mu \nu}$ is a linear combination of all the quadratic contractions of the field strength $C^{\mu \nu \alpha}$. There are nearly a hundred of these but, because of the symmetry properties of $C^{\mu \nu \alpha}$, all but two of them are linearly dependent. Thus, we take the Lagrangian density for $\phi^{\mu \nu}$ to be
\begin{equation}
\mathcal{L}_\phi = \frac{1}{2g^2}(C_{\mu \nu \alpha}C^{\mu \nu \alpha}-C^{\alpha}_{\mu \alpha}C^{\mu \beta}_{\beta})
\label{eq:philag}
\end{equation}

\noindent so the action of a matter system interacting with $\phi^{\mu \nu}$ is
\begin{equation}
S = \int (\mathcal{L}_\phi+\mathcal{L}_{matter})d^4x.
\label{eq:act}
\end{equation}

\noindent where $\mathcal{L}_{matter}$ describes whatever particles and/or fields are involved and has all its partial derivatives replaced by $T_4$ gauge covariant derivatives. Actually the combined Lagrangian density is often not $T_4$ invariant, but when it fails to be so it varies only by a $4$-divergence. This means that the action, and the equations of motion, are $T_4$ gauge invariant. To successfully quantize $\phi^{\mu \nu}$ a particular gauge must be chosen and imposed via a gauge fixing term in $\mathcal{L}_\phi$~\cite{Hsu-09}. Since we are not discussing the quantum aspects of YMG in this paper we decline to choose a gauge to make the equations a little bit simpler. Insisting that Eq.~\ref{eq:act} is stationary gives the equations of motion for $\phi^{\mu \nu}$:
\begin{eqnarray}
g^2T^{\mu \nu} &=& \partial_{\lambda}(J^{\lambda}_{\rho}C^{\rho \mu \nu} - J^{\lambda}_{\alpha}C^{\alpha \beta}_{\beta}\eta^{\mu \nu} + C^{\mu \beta}_{\beta}J^{\nu \lambda}) \nonumber \\
&-& C^{\mu \alpha \beta}\partial^{\nu}J_{\alpha \beta} + C^{\mu \beta}_{\beta}\partial^{\nu}J^{\alpha}_{\alpha} - C^{\lambda \beta}_{\beta}\partial^{\nu}J^{\mu}_{\lambda} \label{eq:phimotion}
\end{eqnarray}

\noindent where $T^{\mu \nu}$ contains all the terms involving matter fields from $\mathcal{L}_{matter}$. Now we can see why YMG might be considered a theory of gravity; in the weak field limit Eq.~\ref{eq:phimotion} becomes mathematically identical to the linearized Einstein's field equations~\cite{Hsu-06}. Requiring that we recover Newtonian gravity in this limit fixes the coupling constant $g = \sqrt{8\pi G}$. Formal mathematical similarity to gravitational equations by its self is not enough to be sure that YMG describes the gravitational interaction. We will now show with an example that interaction with the field $\phi^{\mu \nu}$ modifies the trajectories of objects just like they were in a gravitational field.\newline
\indent If $\mathcal{L}_{matter}$ is the usual Lagrangian density for electrodynamics with $T_4$ gauge covariant derivatives then the equations of motion obtained by varying Eq.~\ref{eq:act} with respect to the electromagnetic potential $A^{\mu}$ are
\begin{equation}
(J_{\mu \nu}\partial^{\nu}+(\partial_{\lambda}J^{\lambda}_{\mu}))(D^{\mu}A^{\alpha}-D^{\alpha}A^{\mu}) = 0.
\label{eq:emmotion}
\end{equation}

\noindent Assume an eikonal form for the potential $A^\mu = a^\mu \exp (iS)$ with $a^\mu$ a constant polarization vector and $S$ the eikonal. Now if we consider the classical limit of these equations (corresponding to ray optics) by taking $|\partial_\mu S| \gg |\phi^{\mu \nu}|$ we obtain
\begin{equation}
G^{\mu \nu}\partial_{\mu}S \partial_{\nu}S = 0,\quad G^{\mu \nu} \equiv J^{\mu \alpha}J_{\alpha}^{\nu}.
\label{eq:efmet}
\end{equation}

\noindent The tensor $G^{\mu \nu}$ is the \emph{effective metric tensor} of YMG. We call it a metric because it modifies the trajectories of classical objects just as a curved geometric metric tensor would. On the other hand it is merely effective because its derivation in no way invoked curved spacetime geometry, and the true geometry of spacetime is taken in this theory to be flat. Although we have only demonstrated that the effective metric appears in the optical limit of electromagnetism, it also appears in the macroscopic-object limit of the Dirac equation for Fermions with $T_4$ covariant derivatives~\cite{Hsu-06}. In every case considered thus far the effective metric tensor, with the same form as the one in Eq.~\ref{eq:emmotion}, appears in the classical limit of equations of motion derived from actions like~\ref{eq:act}. Based on this we see that YMG describes gravity as an emergent phenomena. Microscopic objects interact with it via the full equations of motion, but the cumulative effect of these interactions is to mimic curved spacetime on large scales. We note, for fairness, that no general proof yet exists that the effective metric will always show up for $\phi^{\mu \nu}$ coupled to an arbitrary field.\newline
\indent Recently~\cite{Hsu-13} Hsu has shown that Eq.~\ref{eq:efmet} is not quite right. When $\phi^{\mu \nu}$ couples to another gauge boson field, a choice of gauge must be made before the eikonal form can be used and the classical limit taken. Then summing over the polarizations which $a^{\mu}$ can take in Eq.~\ref{eq:emmotion} leads to a slightly different effective metric tensor:
\begin{equation}
G^{\mu \nu}_L = G^{\mu \nu} - \frac{g}{4}\phi^{\mu}_{\sigma}J^{\sigma \nu}
\label{eq:efmetl}
\end{equation}

\noindent where $G^{\mu \nu}$ is the effective metric from Eq.~\ref{eq:efmet} and the subscript $L$ stands for ``light ray." As discussed in~\citen{Hsu-13} the second term in Eq.~\ref{eq:efmetl} is much smaller than the first so we will neglect it hereafter. Conceptually it is unclear what effect, if any, gauge bosons following a slightly different effective metric than fermions do has on the cosmological model presented in the next section. Investigating these effects, or their absence, is an interesting line of research which has not yet been explored. Thus, we will proceed using Eq.~\ref{eq:efmet} as the definition of the effective metric tensor in YMG.\newline
\indent We conclude this review by noting that YMG seems to have the same predictive power as GR on large scales. Gravitational radiation, periapsis precession, and gravitational lensing are all predicted by YMG and these predictions agree with present observations~\cite{Hsu-11}. In all cases studied so far the results produced are the same as those from GR but with small corrections. This is not too surprising since the theories agree exactly in the weak field limit. The corrections to GR offered by YMG are obscured by present-day experimental uncertainties. They are large enough, however, that one can reasonably expect that experimental precision will someday be able to measure them (\emph{e.g.} they are not Planck-scale corrections).

\section{RW Cosmology in YMG}
The solar-system scale phenomena mentioned above are treated in YMG by solving Eq.~\ref{eq:phimotion} with $T^{00} = m\delta(r)$ and all other components of $T^{\mu \nu}$ equal to zero. The static spherically symmetric solution is not available in closed form as it is for GR but rather has to be found by an approximate ansatz solution. In those cases the solution obtained is a series of inverse powers of $r$ with only the first coefficient chosen to reproduce Newtonian gravity. To start doing cosmology with YMG we take the same basic approach: assume the effective metric has the form that GR's metric takes. In this case that amounts to assuming that $G^{\mu \nu}$ from Eq.~\ref{eq:efmet} is of the Robertson-Walker (RW) form $diag(1,-a^2,-a^2,-a^2)$ where the scale factor $a$ is a function of comoving time alone. Since the effective metric governs the motion of macroscopic objects, the same arguments that lead to the RW form for GR's curved spacetime metric from the assumptions of homogeneity and isotropy will do so for the effective metric as well.\newline

Taking the RW metric as an \emph{ansatz} solution for Eq.~\ref{eq:phimotion} we arrive at YMG's analogue of the Friedmann equation~\cite{katz}
\begin{equation}
H = \frac{H_0}{2a}\sqrt{\Omega_m/a^3 + \Omega_r/a^4 + \Omega_\Lambda}
\label{eq:fried}
\end{equation}

\noindent where $H$ is Hubble's parameter, $H_0$ its value at the present time, $a$ is the scale factor normalized to unity at the present time, and $\Omega_m$, $\Omega_r$, and $\Omega_\Lambda$ are the fractional densities of matter (baryonic and dark), radiation and relativistic particles, and vacuum energy, respectively. Equation~\ref{eq:fried} is the basis for studying cosmology with YMG, just as the Friedmann equations are for GR. Unlike the Friedmann equations, ours has no ingress for a possible curvature term since $K=0$ is a fundamental assumption of YMG. This is not a handicap, though, as observations pretty well rule out $K \neq 0$~\cite{flat1,flat2}. Just as in GR it turns out that there are several special cases for which Eq.~\ref{eq:fried} can be solved exactly. Table~\ref{tab1} shows these solutions up to an overall normalization factor as well as the corresponding standard solutions for comparison.


\begin{table}
\tbl{Scale factors for universes dominated by one type of substance. In each case YMG predicts a slightly slower growing universe than GR does. In the dark energy row, $H$ and $\beta$ are time-independent and proportional to the cosmological constant; see Ref.~\citen{katz}.}{
\begin{tabular}{|l l l|}\hline
	& GR & YMG \Tstrut \\
	Matter & $a \propto t^{2/3}$ & $a \propto t^{2/5}$ \\
	Radiation & $a \propto t^{1/2}$ & $a \propto t^{1/3}$ \\
	Dark Energy $\quad$ & $a \propto e^{Ht}\qquad$ & $a \propto \cosh (\beta t)$ \\ \hline
\end{tabular} \label{tab1}}
\end{table}

\noindent Because $a \sim T^{-1}$ the slower growth as a function of time in YMG corresponds to \emph{faster} growth as a function of temperature. We will see later that this gives some high temperature photons a better chance of escaping reabsorption into the primeval plasma. It is also worth noting that a constant vacuum energy produces more-or-less exponential growth in YMG meaning that the theory should be able to accommodate an early era of extremely rapid inflation due to one or more scalar fields. Inflation in YMG has not yet been studied in detail and in any case is beyond the scope of this article.

\section{Ionization Fractions}
The radiation dominated solution to the Friedmann-like Eq.~\ref{eq:fried} begins as a singularity, so the qualitative aspects of the Big Bang theory are preserved in YMG. Because of this we expect that at some point the hot plasma of mostly protons and electrons will cool sufficiently to allow neutral hydrogen atoms to form and the probability of it scattering with the background photons will plummet. This process happened long ago and thus is not directly observable. However, its influence on the background photon bath is observable by measuring these photons; the CMB. Using the ionization history of the universe to predict anisotropies in the CMB is a complicated problem, and its application in YMG is a work in progress. Of course, a necessary first step is to work out what the ionization history looks like, which is the main topic of this paper. Big Bang Nucleosynthesis (BBN) has not yet been studied in the context of YMG so we begin by assuming that, in accordance with observations, the primordial plasma was $76\%$ protons, $24\%$ helium nuclei, and enough electrons to make it all electrically neutral~\cite{abundance}. The relative abundance of light elements produced during BBN can be predicted by YMG with the help of thermodynamics and nuclear physics. This will be the subject of future work and is another important benchmark for YMG to reach. In addition we take the standard values of $\Omega_b = 0.046$ and $\Omega_m = 0.240$ for the fractional densities of baryonic matter and baryonic plus dark matter, respectively.\newline
\indent To assess and emphasize the importance of helium in the recombination process we will consider two different models: one which ignores helium and one which does not. The first model is based on the classic calculation of Peebles~\cite{peebles}. In it he calculates the photon production rate as a function of frequency, but we will assume that all the photons produced by decays of $2s$ and $2p$ states of hydrogen have the same frequency. In exchange for not having to deal with the Lyman $\alpha$ production rate we must instead calculate the escape probability for $\alpha$'s, that is, the probability that a photon emitted in $2p \rightarrow 1s + \gamma$ will never be reabsorbed by another ground state hydrogen atom~\cite{sobolev}. The result is equivalent to that of Peebles. Although the result is standard~\cite{weinberg} it is worth rederiving here because it uses a multitude of approximations. These approximations are known to be valid in GR but we need to make sure that they remain so in YMG rather than blindly accepting them.\newline
\indent Let $n_B$, $n = 0.76n_B$, $n_e$, $n_{nl}$, and $X_i = n_i/n$ be the number densities of baryons, ionized and neutral hydrogen atoms, free electrons, hydrogen atoms in the $nl$ state, and the fractional abundance of $i = e,\ nl$ relative to hydrogen, respectively. If we assume that the excited states are all in thermal equilibrium then
\begin{equation}
n_{nl} = (2l+1)n_{2s}e^{(E_2-E_n)/kT}
\label{eq:eq}
\end{equation}

\noindent where $E_n$ is the binding energy of the $n^{th}$ excited state and $k$ is Boltzmann's constant. Because the excited states are all in thermal equilibrium with each other de-excitation to the ground state from states having $n>2$ is very inefficient. Thus, in addition to the ground state $1s$ we need only include the $2s$ and $2p$ states in our calculation. The number density of free electrons is reduced by the recombination of hydrogen atoms at a rate proportional to both components; free electrons and free protons. To have charge neutrality means $n_e = n_p$. Likewise, free electrons are produced by photons from $2s \rightarrow 1s + \gamma + \gamma$ transitions at a rate proportional to $n_{2s}$ (because, as we have stated, higher order transitions are inefficient). Symbolically this reads
\begin{equation}
\frac{d}{dt}\left(\frac{n_e}{n}\right) = -\alpha \frac{n_e^2}{n} + \beta \frac{n_{2s}}{n}
\label{eq:dne}
\end{equation}

\noindent where $\alpha$ is the case B recombination coefficient and $\beta$ the reionization coefficient. Both $\alpha$ and $\beta$ are functions of temperature alone and are related to each other by the equilibrium condition eqn.~\ref{eq:eq}
\begin{equation}
\frac{\beta}{\alpha} = \left(\frac{m_ekT}{2\pi}\right)^{3/2}e^{-E_2/kT}
\end{equation}

\noindent where $m_e$ is the mass of the electron. Again because the excited states are in equilibrium increases in the free electron density are enhanced by decays from $n=2$ states and impeded by excitations from ground to $n=2$ states. Thus,
\begin{equation}
\alpha n_e^2 - \beta n_{2s} = (\Gamma_{2s}+3P\Gamma_{2p})n_{2s}-\Gamma_{1s}n_{1s}
\label{eq:n1}
\end{equation}

\noindent where $\Gamma_{1s},\ \Gamma_{2s},$ and $\Gamma_{2p}$ are the rates of $1s+\gamma +\gamma \rightarrow 2s\ \text{or}\ 2p$, $2s \rightarrow 1s + \gamma + \gamma$, and $2p \rightarrow 1s + \gamma$, respectively. We neglect the two-photon decay of the $2p$ state because it is much slower than that of the $2s$ state~\cite{crc}. In the above equation $P$ is the probability that a $2p\rightarrow 1s + \gamma$ photon will escape the plasma without exciting a neighbouring atom, which we will calculate soon. We will see that recombination begins at a higher temperature here than it does in the standard model, but this temperature is still quite a bit smaller than $(E_2-E_3)/k \approx 22,000\ K$. To a good approximation then, we can write
\begin{equation}
n = n_{1s}+4n_{2s}.
\label{eq:n2}
\end{equation}

\noindent In equilibrium the right-hand-side of Eq.~\ref{eq:n1} vanishes and this yields a relationship between all of our decay/excitation rates:
\begin{equation}
\frac{\Gamma_{1s}}{\Gamma_{2s}+3P\Gamma_{2p}} = e^{(E_2-E1)/kT}.
\label{eq:n3}
\end{equation}

\noindent The purpose of deriving the last three equations was to be able to write Eq.~\ref{eq:dne} as a differential equation of a single unknown function. All that remains is to calculate the photon escape probability. Without making any further approximations this is
\begin{equation}
P(t) = \int_{-\infty}^{\infty} d\omega \mathcal{P}(\omega)\exp \left[ -\int_t^{\infty}dt'n_{1s}(t')\left(\frac{3}{2}\right)\left(\frac{2\pi^2\Gamma_{2p}}{k_\alpha^2}\right)\mathcal{P}\left(\omega \frac{a(t)}{a(t')}\right)\right]
\label{eq:P}
\end{equation}

\noindent where $k_\alpha = (E_1-E_2)$ is the Lyman $\alpha$ wavenumber, and
\begin{equation}
\mathcal{P}(\omega) = \frac{\Gamma_{2p}}{2\pi}\frac{1}{(\omega-\omega_\alpha)^2+\Gamma_{2p}^2/4}
\end{equation}

\noindent ($\omega_\alpha$ is the frequency corresponding to $k_\alpha$) is the Breit-Wigner formula~\cite{bw} for the photon-hydrogen resonant cross-section. It would be quite difficult to carry out the integrals in~\ref{eq:P} as they stand, but several approximations can be made with only a minimal loss of precision. First, because recombination happened in the distant past the time scale on which a photon may be absorbed is much less than $1/H_0$ so we can change the argument of $n_{1s}$ from $t'$ to $t$. For the same reason we can Taylor expand $a(t)/a(t') \approx 1-H(t)(t'-t)$. After changing variables from $t'$ to $\omega' = (1-H(t)(t'-t))\omega$ the escape probability is
\begin{equation}
P(t) = \int_{-\infty}^\infty d\omega \mathcal{P}(\omega)\exp \left[-\frac{3\pi^2\Gamma_{2p}n_{1s}}{\omega k_\alpha^2H}\int_{-\infty}^\omega \mathcal{P}(\omega')d\omega'\right].
\end{equation}

\noindent The probability density $\mathcal{P}$ is nearly a delta function of $\omega_\alpha$ so we can replace the factor of $1/\omega$ with $1/\omega_\alpha$. Now all the integrals can be evaluated exactly, giving
\begin{equation}
P(t) = \frac{1-e^{-x}}{x},\quad x = \frac{3\pi^2\Gamma_{2p}n_{1s}}{\omega_\alpha^3H}.
\end{equation}

\noindent Plugging in numbers we see that $x$ is large over our whole range of interest and so we can drop the decaying exponential in $P(t)$. Rearranging the result reveals that we didn't need to know $\Gamma_{2p}$ after all;
\begin{equation}
3P\Gamma_{2p} = \frac{8\pi H\omega_\alpha^3}{n_{1s}}
\label{eq:n4}
\end{equation}

\begin{figure}[h!]
\begin{center}
\includegraphics[scale=0.7]{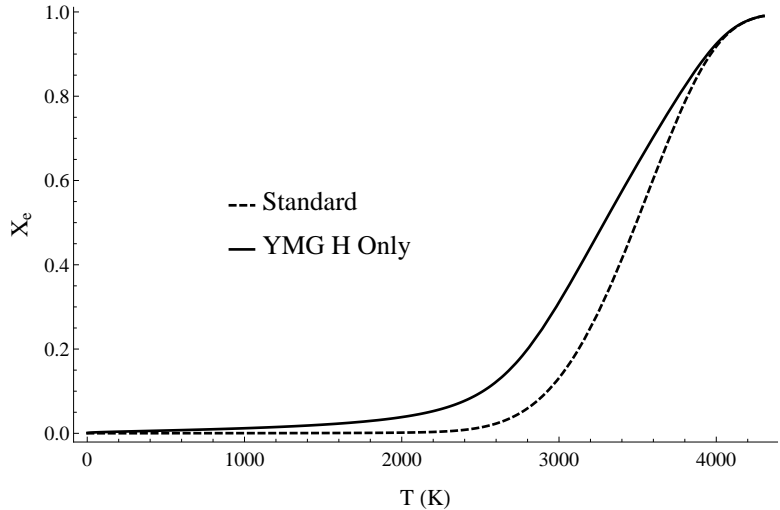}
\end{center}
\caption{Fractional free electron densities ($X_e=n_e/n$) for YMG with hydrogen only and for standard cosmology. The standard cosmology result was also calculated by assuming a hydrogen-only plasma, but in that case including helium makes virtually no difference on the scale of this figure.}
\label{fig:xe}
\end{figure}

\noindent At last, we can combine eqns~\ref{eq:n1},~\ref{eq:n2},~\ref{eq:n3}, and ~\ref{eq:n4} to put Eq.~\ref{eq:dne} in the desired form. For convenience we replace time with temperature as the independent variable, and for simplicity we drop several decaying exponentials which are negligible for the temperatures we're interested in. The result is
\begin{equation}
\frac{dX_e}{dT} = \frac{\alpha n}{HT}\left(1+\frac{\beta}{\Gamma_{2s}+8\pi H\omega_\alpha^3/n(1-X_e)}\right)^{-1}\left(X_e^2-\frac{1-X_e}{S}\right)
\label{eq:dx}
\end{equation}

\noindent where $S = n(m_ekT/2\pi)^{-3/2}\exp (E_1/kT)$ is so named because it is the function which appears in the Saha equation for a similar non-expanding hydrogen plasma. Because of the relationship between $\alpha$ and $\beta$ we only need to specify one of them. P$\acute{e}$quignot, Petijean, and Boisson~\cite{alpha} have compiled a large amount of atomic data and found that 
\begin{equation}
\alpha (T) = F_H\frac{aT^b}{1+cT^d}
\label{eq:ah}
\end{equation}

\noindent fits very well over a wide range of temperatures, both far below and far above our region of interest. (The values of the parameters $F_H,a,b,c$, and $d$ are given in table~\ref{tab2}). With all of this at hand we are ready to solve Eq.~\ref{eq:dx} numerically, using the Saha equilibrium value of $X_e$ as the initial condition. This is yet another approximation, since the hydrogen plasma can't be in Saha equilibrium while it's in an expanding universe, but both values (the Saha and true values) are so close to unity at our starting temperature that it hardly matters. Figure~\ref{fig:xe} shows the numerical solution to Eq.~\ref{eq:dx} along with the standard result from a GR based treatment.\newline
\indent As might have been expected from the slower expansion rates (see table~\ref{tab1}) of YMG, the recombination of free electrons and protons into neutral hydrogen atoms is somewhat delayed. In the figure we have to specify that the blue curve is for the hydrogen-only model in YMG because later on we will consider other models. No such distinction is necessary for the standard result because the influence of helium on the recombination temperature is small enough to not be easily noticeable on a linear-scale graph over the given range.\newline
\indent The qualitative behaviour of the YMG ionization fraction curve provides a useful consistency check for our work so far. The YMG and GR curves more or less coincide for $T \lesssim 1000\ K$ and for $T \gtrsim 4000\ K$, taking on approximate null and unity values, respectively. The high-temperature side of this means that the two cosmologies both have an early universe in which neutral atoms cannot form without being immediately ionized. The low-temperature side of this coincidence is more interesting. In both theories the tendency for free electrons and protons to combine into neutral hydrogen eventually overwhelms both the ionization (due to ambient
\newpage

\noindent radiation) rate and cosmological expansion. This is an old result for the standard cosmology but it is noteworthy for YMG because it is not obvious that Eq.~\ref{eq:fried} allows for such an interplay of rates. As mentioned above, the leisurely expansion rates in table~\ref{tab1} delay the recombination process and produce disagreement between GR and YMG for $1000\ K\ \lesssim T \lesssim 4000\ K$. It remains to be seen how this disagreement manifests in the predicted CMB power spectrum; the cosmological perturbation theory of YMG is still a work in progress.\newline
\indent Having seen the timeline for recombination without helium we are ready to appreciate its importance. To maximize the level of detail while minimizing complexity we use the effective three-level model of Seager \emph{et al}~\cite{seager,seager1} based on their enormous $300$-level simulation. To reduce their system of thousands of ODE's to just three, they absorb a lot of information into the effective recombination and reionization coefficients $\alpha$ and $\beta$. Of course, in our model, the Hubble parameter follows a different evolution equation but the thermal physics is the same. The ionic fractions $X_H,\ X_{He}$ (hydrogen and helium II) and $T_M$, the temperature of non-relativistic matter, are the dependent variables to be solved for while $T$, the temperature of the radiation background, is the sole independent variable. We also take $T$, rather than $T_M$, as the driving force behind cosmic expansion, that is, $H = H(T)$. Using $X_e=0.76X_H+0.24X_{He}$ the equations coupling $X_H,\ X_{He}$, and $T_M$ are
\begin{eqnarray}
\frac{dX_H}{dT} &=& \left(X_eX_Hn_H\alpha_H-\beta_H(1-X_H)e^{-\omega_{H2s}/kT_M}\right) \nonumber \\
&\times& \frac{1+K_H\Gamma_{H2s}n_H(1-X_H)}{HT(1+K_H(\Gamma_{H2s}+\beta_H)n_H(1-X_H))} \nonumber \\
\frac{dX_{He}}{dT} &=& \left(X_{He}X_en_H\alpha_{He}-\beta_{He}(f_{He}-X_{He})e^{-\omega_{He2s}/kT_M}\right) \label{eq:dxhe} \nonumber \\
&\times& \frac{1+K_{He}\Gamma_{He2s}n_H(f_{He}-X_{He})e^{-\omega_{ps}/kT_M}}{HT(1+K_{He}(\Gamma_{He2s}+\beta_{He})n_H(f_{He}-X_{He})e^{-\omega_{ps}/kT_M})} \nonumber \\
\frac{dT_M}{dT} &=& \frac{8\sigma_Ta_RT^3}{3Hm_e}\frac{X_e}{1+f_{He}+X_e}(T_M-T)+\frac{2T_M}{T}
\end{eqnarray}

\noindent Here $\alpha_{He}$ is an effective recombination coefficient which is fitted to incorporate the data from Seager's $300$-level model. It takes its form from Verner and Ferland~\cite{verner} and is most valid between $4000$ and $10,000\ K$ but it is still very accurate down to $1000\ K$.
\begin{equation}
\alpha_{He} = F_{He}q\left[\sqrt{\frac{T_M}{T_2}}\left(1+\sqrt{\frac{T_M}{T2}}\right)^{1-p}\left(1+\sqrt{\frac{T_M}{T1}}\right)^{1+p}\right]^{-1}
\label{eq:ahe}
\end{equation}

\noindent The functions $K_H$ and $K_{He}$ are shorthand for the cosmologically redshifted wavelengths of $2p \rightarrow 1s$ photons from hydrogen and helium I, respectively: $K_i = (8\pi \omega_{i2p} H)^{-1}$, $i = H,\ He$. Table~\ref{tab2} summarizes the constants and parameters in Eqs.~\ref{eq:ah},~\ref{eq:dxhe}, and~\ref{eq:ahe}. As recommended by Seager \emph{et al} the parameters $F_H$ and $F_{He}$ are order unity fudge factors which absorb some of the error introduced in approximating a $300$-level model by a $3$-level model.\newline

\renewcommand{\arraystretch}{1.2}
\begin{table}
\tbl{Definitions and values of the constants and parameters used in the recombination simulations. $1)$ Strictly speaking it doesn't make sense to call $\omega_{2s}$ the frequency of \emph{the} photon in a two-photon $2s \rightarrow 1s$ transition. Rather, these frequencies correspond to the frequency of a hypothetical photon carrying all the energy of the transition. The total energies of transitions are the quantities which appear in our equations, so we don't need to worry about how that energy is shared between the photons in two-photon decays. Still, we note in passing that the equipartition of energy amongst the photons is the most likely decay mode of a hydrogen-like atom~\cite{2photon}. $2)$ To clarify, $f_{He} = Y_p/4(1-Y_P)$ where $Y_p=0.24$ is the primordial helium abundance.}{
\begin{tabular}{|c|c|c|}\hline
\textbf{Symbol} & \textbf{Value} & \textbf{Meaning} \\ \hline \hline
$\Gamma_{H2s}$ & $8.22458\ s^{-1}$ & $H\ 2s \rightarrow 1s$ two-photon decay rate \\ \hline
$\Gamma_{He2s}$ & $51.3\ s^{-1}$ & $HeI\ 2s \rightarrow 1s$ two-photon decay rate \\ \hline
$\omega_{H2s}$ & $1.5505\times 10^{16}\ s^{-1}$ & $H\ 2s \rightarrow 1s$ photon frequency$^1$\\ \hline
$\omega_{He2p}$ & $3.2258\times 10^{16}\ s^{-1}$ & $HeI\ 2p \rightarrow 1s$ photon frequency \\ \hline
$\omega_{He2s}$ & $3.1343\times 10^{16}\ s^{-1}$ & $HeI\ 2s \rightarrow 1s$ photon frequency$^1$ \\ \hline
$\omega_{ps}$ & $9.1559\times 10^{14}\ s^{-1}$ & $\omega_{He2p}-\omega_{He2s}$ \\ \hline
$a$ & $4.309\ m^3 s^{-1}$ & \multirow{5}{*}{Parameters for hydrogen recombination coefficient} \\ \cline{1-2}
$b$ & $-0.6166$ & \\ \cline{1-2}
$c$ & $0.6703$ & \\ \cline{1-2}
$d$ & $0.5300$ & \\ \cline{1-2}
$F_H$ & $1.14$ & \\ \hline
$q$ & $10^{-16.744}\ m^3s^{-1}$ & \multirow{5}{*}{Parameters for helium I recombination coefficient} \\ \cline{1-2}
$p$ & $0.711$ & \\ \cline{1-2}
$T_1$ & $10^{5.114}\ K$ & \\ \cline{1-2}
$T_2$ & $3\ K$ & \\ \cline{1-2}
$F_{He}$ & $3$ & \\ \hline
$f_{He}$ & $0.0789$ & Number fraction of $He$ to $H\ ^2$ \\ \hline
$a_r$ & $4722.2\times 10 ^{-6}\ eV cm^{-3}K^{-4}$ & Radiation constant \\ \hline
$\sigma_T$ & $6.6525\times 10^{-23}\ cm^2$ & Thomson scattering cross-section \\ \hline
\end{tabular}
\label{tab2}}
\end{table}

\indent Solving Eqs.~\ref{eq:dxhe} yields the three functions $X_H,\ X_{He},$ and $T_M$ of the radiation temperature, but we're only interested in a certain linear combination of two of them: $X_e = 0.76X_H+0.24X_{He}$. This function, the fraction of free electrons, is plotted below in figure~\ref{fig:xehe}.
\begin{figure}[h!]
\begin{center}
\includegraphics[scale=0.7]{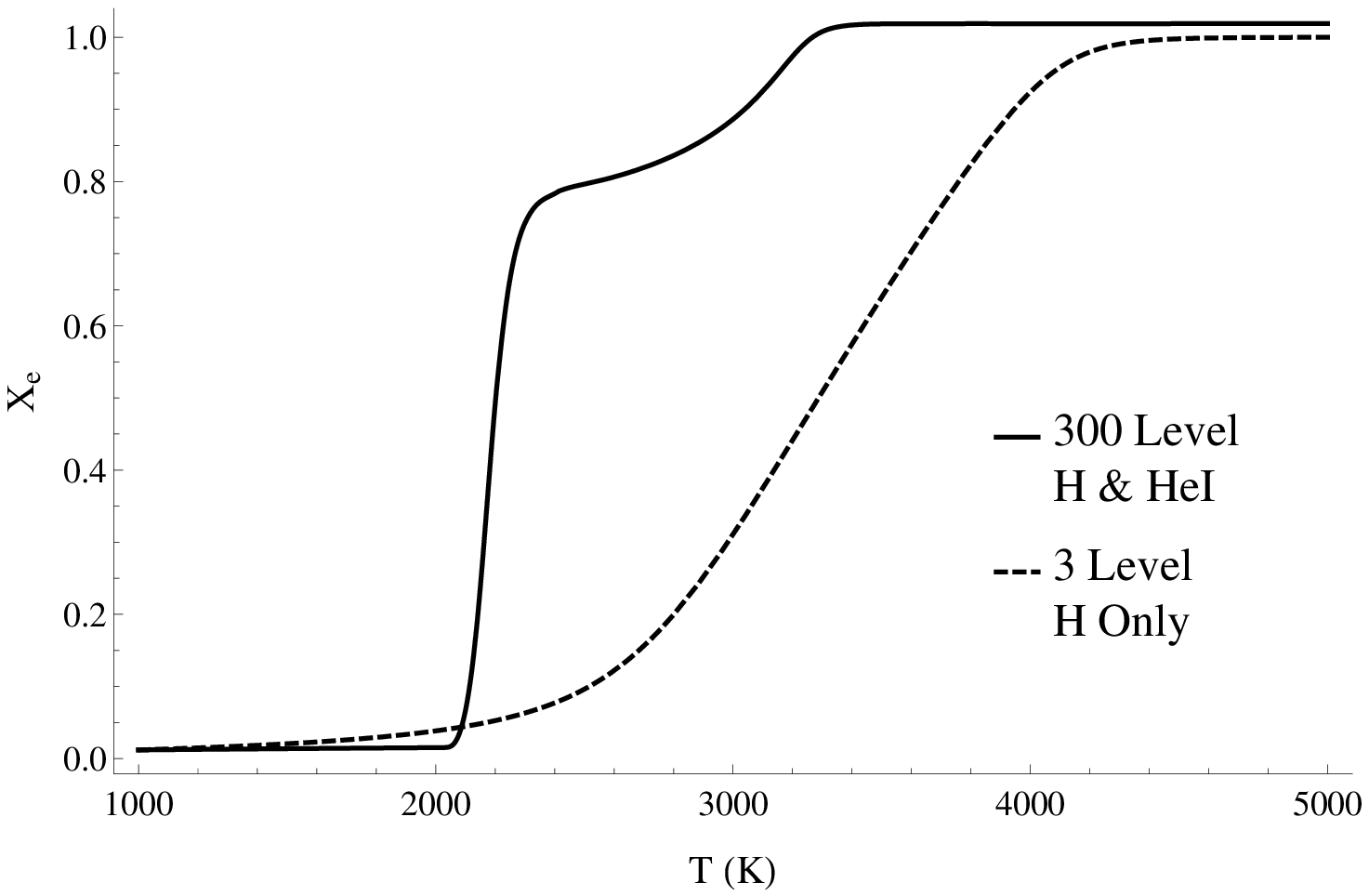}
\end{center}
\caption{The fraction of free electrons, as a function of radiation temperature, according to YMG. This figure shows that the inclusion of helium into the calculation makes a substantial difference in the ionization history of the early universe. The first (highest temperature) drop in the $H\ \&\ He$ model is the recombination of $HeII$ to $HeI$ and the second drop is the recombination of hydrogen.}
\label{fig:xehe}
\end{figure}

\noindent The fact that accounting for singly-ionized helium changes the recombination rate so drastically tells us that, unlike in the standard model, the hydrogen and helium recombinations overlap significantly. Qualitatively this makes sense because the expansion rate of the universe is slower in YMG than it is in GR. According to $a\sim T^{-1}$ expanding slower as a function of time means expanding faster as a function of temperature. Thus, at the (high) temperatures at which helium would normally finish recombining the expansion of the universe makes it more diffuse and so slows down the interaction rate. This means that both protons ($H^+$) and $HeII$ are competing for the same cloud of electrons over a wider range of temperatures. Once all the helium is done recombining, at around $2100\ K$, the remaining $H^+$ quickly captures the rest of the free electrons.\newline
\indent The details of how the universe transitioned from fully ionized to electrically neutral on large scales are different in YMG than they are in GR, though qualitatively they are similar. In both cases recombination is significantly delayed, compared to static Saha equilibrium, by cosmic expansion. Also there is a specific temperature at which virtually all of the recombination occurs. The theories differ in their prediction of what that temperature is and how long the transition takes. As we have stated before, the ionization history its self is not directly observable (we've missed it by billions of years) but it, along with a gravitational theory of small fluctuations in an expanding universe, leads to the power spectrum of the CMB. This prediction requires that we know the probability, as a function of temperature, that a given photon last scattered at that temperature before eventually being absorbed by one of our detectors: the visibility function.

\section{Opacity \& Visibility Functions}
The optical depth of the hydrogen/helium plasma at some time $t$ is
\begin{equation}
\tau(t) = -\int_t^{t_0}\sigma_Tn_e(t')dt'
\label{eq:tau}
\end{equation}

\noindent where $n_e$ is the number density of free electrons and $t_0$ is the present time. Several reasonable approximations go into this definition. First, by using the Thomson cross-section we are ignoring relativistic collisions between electrons and photons. Over the temperature range on which we plan to use Eq.~\ref{eq:tau}, $\sim 1000\ K$ to $\sim 5000\ K$ this is thoroughly acceptable. Secondly, since only $n_e$ appears in Eq.~\ref{eq:tau}, we are neglecting the possibility of photons scattering off of neutral hydrogen and/or helium atoms. This is also fine because the probability of a photon scattering off an electrically neutral particle is dwarfed by the probability of it scattering off a charged particle. We have calculated $n_e$ as a function of temperature so it will be more convenient to change the variables of the integral in $\tau$ from $t$ to $T$, giving
\begin{equation}
\tau(T) = -\sigma_T\int_{T_0}^T \frac{n_e(T')}{H(T')T'}dT'
\label{eq:tauT}
\end{equation}

\noindent where $T_0 = 2.725\ K$ is the present temperature of the CMB. What we really want to do is identify the temperature of last scattering, and for that we invoke the visibility function~\cite{rybicki}
\begin{equation}
F(T) = \left(\frac{d}{dT'}e^{-\tau(T')}\right)\left . \right |_{T'=T}.
\end{equation}

\begin{figure}[h!]
\begin{center}
\includegraphics[scale=0.7]{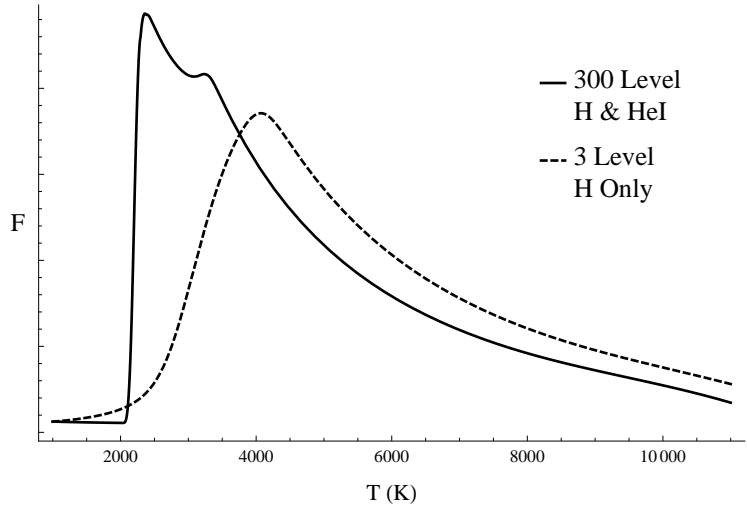}
\end{center}
\caption{Visibility functions, in arbitrary units, versus temperature for the two models.}
\label{fig:vis}
\end{figure}

\noindent In the standard picture $F$ is very well fit by a Gaussian probability distribution with mean $2941\ K$ and spread $\sigma = 248\ K$~\cite{weinberg}. The same cannot be said for our visibility function, which is shown in figure~\ref{fig:vis} for the hydrogen-only model and for the multilevel hydrogen \& helium model.

\noindent Again we see that the inclusion of helium in the calculations makes a huge difference. Not only does it push the visibility function's peak to a lower temperature, it also makes the cut-off much sharper. The standard model result is not shown in figure~\ref{fig:vis} because it is much larger on the vertical scale and much smaller on the horizontal than the YMG visibility functions. Nevertheless we remark that it too has a high-temperature tail which decays slower than the one on the other side of the peak. The fact that our visibility functions have a cut-off means that YMG does predict that there is a temperature at which matter and radiation essentially decouple. We can understand the relatively high probability of last scattering at higher temperatures by recalling that the scale factor increases at a faster rate (compared to that of GR) at high temperatures. Some of the photons scattered at, say, $7000\ K$ will be left stranded by this rapid expansion and never hit another atom (until we measure it).

\section{Conclusions and Future Work}
The discrepancy between the ionization history according to YMG and GR might be a cause for concern. After all, GR has been used very successfully to predict the power spectrum of the CMB and other cosmological features. Using the visibility function in figure~\ref{fig:vis} with GR's perturbation theory would necessarily lead to results contrary to observation. But YMG is a theory of gravity independent of GR and as such, deviations from perfect homogeneity and isotropy must be dealt with by perturbing the energy-momentum and RW effective metric tensors. Equation~\ref{eq:phimotion} then gives us the means to deduce the evolution of these perturbations. We have derived these perturbed equations and will use them, along with the usual theory of thermodynamics in an expanding universe, to calculate the power spectrum of the CMB, for comparison with observations. Conceptually this calculation will be very much like that of GR. Technically, because of the large spread of the visibility function in YMG, we will be unable to make use of the ``sudden decoupling" approximation.\newline
\indent We have shown that YMG predicts an era of recombination of neutral atoms from the primordial plasma. This process takes much longer in YMG than it does in GR, but both theories have a sharp cut-off in their cosmic visibility functions. Including helium in the recombination calculations of GR induces only very small corrections whereas it radically changes the dynamics in YMG. We understand this change as being due to a larger overlap between the eras of helium and hydrogen recombination. Our results are not directly observable, but they are required for calculating quantities which are. Future work will use these results to calculate the power spectrum of the CMB in order to test the validity of YMG.

\section*{Acknowledgements}
The author wishes to thank Christopher Roberts for useful discussions on numerical aspects of this project.

\bibliographystyle{unsrt} 
\bibliography{ymbib}
\end{document}